\documentclass[12pt]{article}
\newcommand{\nn}{\nonumber}
\newcommand{\beqn}{\begin{eqnarray}}
\newcommand{\eeqn}{\end{eqnarray}}
\newcommand{\be}{\begin{equation}}
\newcommand{\ee}{\end{equation}}
\newcommand{\eqn}[1]{(\ref{#1})}
\newcommand{\bel}[1]{\be\label{#1}}
\newcommand{\ba}{\begin{array}{c}}
\newcommand{\bat}{\begin{array}{cc}}
\newcommand{\ea}{\end{array}}

\newcommand{\bi}{\begin{itemize}}
\newcommand{\ei}{\end{itemize}}
\newcommand{\chpt}{$\chi$PT}
\newcommand{\rcht}{R$\chi$T}

\newcommand{\no}{\nonumber}

\newcommand{\imag}{{\rm Im}}
\newcommand{\cO}{{\cal O}}

\begin{document}

\begin{titlepage}

\vspace*{-1.8cm}
\begin{flushright}
{\small\sf  IFIC/06$-$36\\ FTUV/06$-$2310 \\}
\end{flushright}

\vspace*{2.2cm} 
\begin{center}
{\Large\bf Towards a determination of the chiral \\ \vspace{0.3cm}

 couplings at NLO in $\mathbf{1/N_C}$: $\mathbf{L_8^r(\mu)}$ and $\mathbf{C_{38}^r(\mu)}$}  

\vspace{0.7cm}

{\normalsize\bf \sc I. Rosell$^{1}$, J.J. Sanz-Cillero$^2$}
and {\normalsize\bf \sc A. Pich$^1$} \\

\vspace{1.2cm} 

${}^{1)}$ {\em Departament de F\'{\i}sica Te\`orica, IFIC, CSIC --- 
Universitat de Val\`encia \\ 
Edifici d'Instituts de Paterna, Apt. Correus 22085, E-46071 
Val\`encia, Spain} \\[10pt]
${}^{2)}$ {\em Department of Physics, Peking University \\ Beijing 100871, P.R. China} \\
\end{center}

\vspace{2.0cm}

\begin{abstract}
We present a dispersive method which allows to investigate
the low-energy couplings of chiral perturbation theory at the
next-to-leading order (NLO) in the $1/N_C$ expansion, keeping full
control of their renormalization scale dependence. Using the
resonance chiral theory Lagrangian, we perform a NLO calculation of
the scalar and pseudoscalar two-point functions, within the
single-resonance approximation. Imposing the correct QCD
short-distance constraints, one determines their difference
$\Pi(t)\equiv\Pi_S(t)-\Pi_P(t)$ in terms of the pion decay constant
and resonance masses. Its low momentum expansion fixes then the
low-energy chiral couplings $L_8$ and $C_{38}$. At $\mu_0=0.77$~GeV,
we obtain $L_8^r(\mu_0)^{SU(3)} = (0.6\pm 0.4)\cdot 10^{-3}$ and
$C_{38}^r(\mu_0)^{SU(3)}=(2\pm 6)\cdot 10^{-6}$.
\end{abstract}

\end{titlepage}

\section{The large--$\mathbf{N_C}$ limit}

In recent years we have witnessed a spectacular progress in our
understanding of low-energy effective field theories \cite{EFT,MA:96,GE:93}. In
particular, chiral perturbation theory (\chpt) has been established
as a very powerful tool to incorporate the chiral symmetry
constraints when analysing the strong interactions in the
non-perturbative regime \cite{GL:85,PI:95,Ec:95}. The precision
required in present phenomenological applications makes necessary to
include corrections of $\cO(p^6)$. While many two-loop \chpt\
calculations have been already performed~\cite{ChPT2loops,ChPT2loopsb,ChPT2loopsc,ChPT2loopsd,ChPT2loopsdbis,ChPT2loopse,ChPT2loopsf,ChPT2loopsg,ChPT2loopsh}, the large number of
unknown low-energy couplings (LECs) appearing at this order puts a
clear limit to the achievable accuracy \cite{PI:05}.

The limit of an infinite number of quark colours has proved very
useful to bridge the gap between \chpt\ and the underlying QCD
dynamics \cite{PI:02,MHA}. Assuming confinement, the strong dynamics
at $N_C\to\infty$ is given by tree diagrams with infinite sums of
hadron exchanges, which correspond to the tree approximation of some
local effective Lagrangian \cite{tHO:74,WI:79}. Resonance chiral
theory (\rcht) provides a correct framework to incorporate the
massive mesonic states within an effective Lagrangian formalism
\cite{RChTa}. Integrating out the heavy fields one recovers the
\chpt\ Lagrangian with explicit values of the chiral LECs in terms
of resonance parameters. Moreover, the short-distance properties of
QCD impose stringent constraints on the low-energy parameters
\cite{RChTb}.

Truncating the infinite tower of meson resonances to the lowest
states with $0^{-+}$, $0^{++}$, $1^{--}$ and $1^{++}$ quantum
numbers (single-resonance approximation, SRA), one gets a very
successful prediction of the $\cO(p^4 N_C)$ \chpt\ couplings in
terms of only three parameters: $M_V$, $M_S$ and the pion decay
constant $F$ \cite{PI:02}. Some $\cO(p^6)$ LECs have been already
predicted in this way, by studying an appropriate set of three-point
functions \cite{three}. More recently, the program to determine
all $\cO(p^6)$ LECs at leading order in $1/N_C$ has been put on very
solid grounds, with a complete classification of the needed terms in
the \rcht\ Lagrangian \cite{CEEKPP:06}.

Since chiral loop corrections are of next-to-leading order (NLO) in
the $1/N_C$ expansion, the large--$N_C$ determination of the LECs is
unable to control their renormalization-scale dependence. For
couplings related with the scalar sector this introduces large
uncertainties, because their $\mu$ dependence is very sizable. A
first analysis of resonance loop contributions to the running of
$L_{10}^r(\mu)$ was attempted in Ref.~\cite{CP:02}. More recently, a
NLO determination of the \chpt\ coupling $L_9^r(\mu)$ has been
achieved, through a one-loop calculation of the vector form factor
in \rcht\ \cite{RSP:05}. In spite of all the complexity associated
with the still not so well understood renormalization of \rcht\
\cite{RSP:05,RPP:05}, this calculation has shown the potential
predictability at the NLO in $1/N_C$.

In this article we present a NLO determination of the couplings
$L_8^r(\mu)$ and $C_{38}^r(\mu)$. Using analyticity and unitarity we can avoid all
technicalities associated with the renormalization procedure,
reducing the calculation to tree-level diagrams plus dispersion
relations. This allows to understand the underlying physics in a
much more transparent way. In particular, the subtle cancellations
among many unknown renormalized couplings found in
Ref.~\cite{RSP:05} and the relative simplicity of the final result
can be better understood in terms of the imposed short-distance
constraints.

Let us consider the two-point correlation functions of two scalar or
pseudoscalar currents, in the chiral limit. Of particular interest
is their difference $\Pi(t)\equiv \Pi_{S}(t)-\Pi_{P}(t)$, which is
identically zero in QCD perturbation theory. When $t\to\infty$, this
correlator vanishes as $1/t^2$, with a coefficient proportional to
$\alpha_s\langle\bar q\Gamma q\,\bar q\Gamma q\rangle$
\cite{SVZ,sspp}. The low-momentum expansion of $\Pi(t)$ is
determined by \chpt\ to have the form \cite{GL:85,ChPT2loops}
\begin{eqnarray}\label{eq:Pi_chpt}
  \Pi(t) &=&  B_0^2 \, \left\{ \frac{2 F^2}{t} \, +\, 32 L_8^r(\mu)
\, +\, \frac{\Gamma_8}{\pi^2} \left( 1-\ln \frac{-t}{\mu^2} \right)\no\right. \\ &&
\quad +\, 
\frac{t}{F^2} \bigg[ 32\,C_{38}^r (\mu)  -\frac{\Gamma_{38}^{(L)}}{\pi^2}  \left( 1-\ln \frac{-t}{\mu^2} \right) +\cO\left(N_C^0\right) \bigg]\,+\,
 \cO\left(t^2\right)\Bigg\}\, ,\label{eq:Pi_chpt}
\end{eqnarray}
with $\Gamma_8 = 5/48$ [$3/16$] and $\Gamma_{38}^{(L)}=-5L_5^r/6$ [$-3L_5^r/2$] in the $SU(3)_L\otimes SU(3)_R$
[$U(3)_L\otimes U(3)_R$] effective theory~\cite{GL:85,ChPT2loops}. 
The correlator is proportional to $B_0^2\equiv \langle \bar q
q\rangle^2/F^4$, which guarantees the right dependence with the QCD
renormalization scale. The couplings $F^2$, $L_8$ and $C_{38}/F^2$
are $\cO(N_C)$, while $\Gamma_8$ and $\Gamma_{38}^{(L)}/F^2$ are
$\cO(1)$ and represent a NLO effect.\footnote{Note that we have
defined a dimensionless $C_{38}$ and $C_{38}^r$; in the notation of
Ref.~\cite{ChPT2loops} that corresponds to $F^2 C_{38}$ and
$C_{38}^r$ respectively.}

In the large--$N_C$ limit, $\Pi(t)$ has the general form
\bel{eq:SSR}
 \Pi(t) \,=\, 2\, B_0^2\,\left\{ \sum_i\,
 {8\, c_{m_i}^2\over M_{S_i}^2-t} \, -\,\sum_i\, {8\, d_{m_i}^2\over
 M_{P_i}^2-t} \, +\, {F^2\over t}\right\} \, ,
\ee
which involves an infinite number of scalar and pseudoscalar meson
exchanges. This expression can be easily obtained within R$\chi$T,
with $c_{m_i}$ and $d_{m_i}$ being the relevant meson couplings. For
a finite number of resonances,\footnote{ Some issues related to the
truncation of the spectrum to a finite number of resonances are
discussed in Refs.~\cite{ximo,Sanz-Cillero:2005ef,tall}.} one finds
that imposing the right high-energy behaviour ($\sim 1/t^2$)
constrains the resonance parameters to obey the
relations~\cite{Golterman:1999au}: \bel{eq:SD5}
 \sum_i\,\left( c_{m_i}^2 \,-\, d_{m_i}^2\right) \,=\, \frac{F^2}{8} \, ,
\quad \quad  \quad\;\; \sum_i\,\left( c_{m_i}^2 \,M_{S_i}^2\, -\, d_{m_i}^2 \,M_{P_i}^2\right)
 \,=\,\widetilde{\delta} \,,
\ee
where $\widetilde{\delta} \equiv 3\,\pi\alpha_s F^4/4\approx 0.08\,\alpha_sF^2 \times
(1~\mathrm{GeV})^2$.
Truncating the infinite sums to their first contributing states and neglecting $\widetilde{\delta}$, these relations fix
the corresponding scalar and pseudoscalar couplings in terms of the
resonance masses:
\bel{eq:cm_cd}
 c_m^2\, =  \, \frac{F^2}{8}\, \frac{M_P^2}{M_P^2-M_S^2} \, ,
\quad \quad \qquad d_m^2\,=\,\frac{F^2}{8}\, \frac{M_S^2}{M_P^2-M_S^2}\, .
\ee
Note that Eqs.~(\ref{eq:cm_cd}) imposes $M_P \geq M_S$.
On the other hand, the low-energy expansion of \eqn{eq:SSR}
determines~\cite{RChTb,CEEKPP:06}
\begin{eqnarray}
L_8 &= &\sum_i\,\left\{ {c_{m_i}^2\over 2\, M_{S_i}^2}\, -\,
 {d_{m_i}^2\over 2\, M_{P_i}^2}\right\}
 \,\approx\, {F^2\over 16\, M_S^2}\,+\,{F^2\over 16\, M_P^2}\, .\no \\
C_{38}&=& \sum_i \, \left\{ \frac{c_m^2 \,F^2}{2M_S^4} -\frac{d_m^2\,F^2}{2M_P^4} \right\} \,\approx \, \frac{F^4}{16 M_P^{2}M_S^{2}} \left( 1 + \frac{M_P^{2}}{M_S^{2}} + \frac{M_S^{2}}{M_P^{2}} \right) \, . \label{eq:L_8_LO}
\end{eqnarray}
%
Using the approximate constraint $M_P/\sqrt{2}\approx M_S \sim
1$~GeV~\cite{JOP:02}, this gives $L_8 \approx 3F^2/(32 M_S^2)\approx
0.7\cdot 10^{-3}$ and $C_{38}\approx 7F^4/(64M_S^4) \approx 7 \cdot
10^{-6}$. However, one does not known at which scale $\mu$ these
predictions apply.

\section{NLO corrections}

At the NLO in $1/N_C$, $\Pi(t)$ has a contribution from one-particle
exchanges, with the structure in Eq.~\eqn{eq:SSR}, plus one-loop
corrections $\Delta\Pi(t)$ generating absorptive contributions from
two-particle exchanges. The corresponding spectral functions of the
scalar and pseudoscalar correlators take the form:
\beqn\label{eq:ImSS}
 \frac{1}{\pi}\,\imag\Pi_{S}(t) & =& 2\, B_0^2\, \left\{ 8\, c_m^2 \,\delta(t-M_S^2)\, +\,
 \frac{3\,\rho_S(t)}{16\pi^2}\right\}\, ,\no
\\
  \frac{1}{\pi}\,\imag\Pi_{P}(t) & =& 2\, B_0^2 \,\left\{ F^2 \,\delta(t)\, +\,
  8\, d_m^2\, \delta(t-M_P^2)\, +\,
  \frac{3\,\rho_P(t)}{16\pi^2}\right\} \,,
\eeqn
with
\beqn
 \rho_S(t) & =& \frac{\theta(t)}{2}\, |F^{\pi\pi}_S(t)|^2 \,+\,
 \theta(t-M_P^2)\, \left(1-\frac{M_P^2}{t}\right) \,|F^{P\pi}_S(t)|^2
\no\\ & +&
 \theta(t-M_A^2)\,
 \frac{t^2}{4M_A^4}\left(1-\frac{M_A^2}{t}\right)^3 \,|F^{A\pi}_S(t)|^2
 \,+ \cdots
\\
 \rho_P(t) & =& \theta(t-M_V^2)\,
 \frac{t^2}{4M_V^4}\,\left(1-\frac{M_V^2}{t}\right)^3 \,|F^{V\pi}_P(t)|^2
\no\\ & +&
 \theta(t-M_S^2) \,\left(1-\frac{M_S^2}{t}\right)\, |F^{S\pi}_P(t)|^2
 \,+ \cdots
\eeqn
We have adopted the single-resonance approximation and, moreover, we
have only taken explicitly into account the lowest-mass two-particle
exchanges: two Goldstone bosons or one Goldstone and one heavy
resonance. In the energy region we are interested in, exchanges of
two heavy resonances or higher multiplicity states are kinematically
suppressed. Our normalization takes into account the different
flavour-exchange possibilities. The relevant two-particle cuts are
governed by the following scalar,
\beqn
F^{\pi\pi}_S(t)& =& \frac{M_S^2}{M_S^2-t}\, ,
\no\\
F^{P\pi}_S(t)& =&
\sqrt{1-\frac{M_S^2}{M_P^2}}\,\frac{M_S M_P}{M_S^2-t}\, ,
\\
F^{A\pi}_S(t)& =& 0\, ,\phantom{\frac{1}{2}}
\no\eeqn
and pseudoscalar,
\beqn
F^{V\pi}_P(t)& =& -2\,\sqrt{1-\frac{M_V^2}{M_A^2}}\,
\frac{M_V^2M_P^2}{(M_P^2-t)t}\, ,
\no\\
F^{S\pi}_P(t)& =& \,\sqrt{1-\frac{M_S^2}{M_P^2}}\,
\frac{M_S^2M_P^2}{(M_P^2-t)t}\, ,
\eeqn
form factors \cite{RSP:06}.

The \rcht\ couplings \cite{RChTa,CEEKPP:06} generating these form
factors have been determined imposing a good high-energy behaviour
of the corresponding spectral functions, i.e. that the individual
form factor contributions to $\rho_S(t)$ and $\rho_P(t)$ should
vanish at infinite momentum transfer. Moreover, we have used the
constraints \eqn{eq:cm_cd} and the analogous relations (Weinberg sum
rules and good high-energy behaviour of the vector form factor)
emerging in the vector/axial-vector sector. It is quite remarkable
that these short-distance constraints completely determine the form
factors in terms of the resonance masses~\cite{RSP:06}. The form
factor $F^{A\pi}_S(t)$ turns out to be identically zero, within the
SRA.\footnote{We are using the SRA and only operators constructed
with up to $\cO(p^2)$ chiral tensors are allowed in the \rcht\
Lagrangian. While the inclusion of extra resonances would certainly
generate a non-zero value for $F^{A\pi}_S(t)$, its numerical impact
in our final resuls would be small, because the short-distance
contraints would force a compensating effect with the other
contributions. The same comment applies to possible
higher-derivative chiral structures, which generate a bad
high-energy behaviour and, therefore, would need to be severely
tuned.}

Using its known analyticity properties, $\Delta\Pi(t)$ can be
obtained from the spectral functions through a dispersion relation,
up to a subtraction term which has the same structure as the
tree-level scalar and pseudoscalar resonance exchanges
. Therefore, the unknown subtraction constants can be
absorbed by a redefinition of $c_m$, $d_m$, $M_S$ and $M_P$ at NLO
in $1/N_C$:
\bel{eq:Pi_SP}
\Pi(t)\, =\, 2\, B_0^2\,\left\{
\frac{8\, c_{m}^{r\, 2}}{M_{S}^{r\, 2} - t}\, -\,
\frac{8\, d_{m}^{r\, 2}}{M_{P}^{r\,  2} - t} \,+\, \frac{F^2}{t}
\,+\, \Delta\Pi(t)|_\rho\right\}\, .
\ee
The explicit expression of $\Delta \Pi (t)|_\rho$ is relegated to appendix~A. 
At large values of $t$, the one-loop contribution has the behaviour
\begin{equation}
\Delta  \Pi(t)|_\rho \,=\,  \frac{F^2}{t}\,
\delta_{_{\rm NLO}}^{(1)} \, +\, \frac{F^2 M_S^2}{t^2} \,\left( \delta_{_{\rm
NLO}}^{(2)}\, +\, \widetilde{\delta}_{_{\rm NLO}}^{(2)} \,
\ln\frac{-t}{M_S^2}\right)\, +\, \cO\left(\frac{1}{t^3}\right)\, .
\end{equation}
Since the logarithmic term $\ln(-t)/t^2$ should vanish, one obtains
the constraint
\begin{equation}
\widetilde{\delta}_{_{\rm NLO}}^{(2)}\,=\, \frac{M_S^2}{32\pi^2 F^2} \left\{ 6 \left( 1-\frac{M_V^2}{M_A^2} \right) \frac{M_P^4}{M_S^4} +3 \left( 1-2\frac{M_P^2}{M_S^2} \right) \right\} \,=\, 0,
\end{equation}
leading to
\begin{equation}\label{log}
\left(1-\frac{M_V^2}{M_A^2}\right) \, =\,
\frac{M_S^2}{M_P^2}\left(1\, - \, \frac{M_S^2}{2\, M_P^2}\right)\, ,
\end{equation}
which requires $M_A \leq \sqrt{2} M_V$. Imposing the right short-distance behaviour ($\sim 1/t^2$) in $\Pi(t)$, one gets
\begin{eqnarray}
F^2\, (1\,+\,\delta_{_{\rm NLO}}^{(1)})\, - \, 8\, c_m^{r\, 2} \, +\, 8 \,
d_m^{r\, 2}\, = \, 0 \, ,  \no
\\
F^2 \,M_S^2 \,\delta_{_{\rm NLO}}^{(2)}\, - \, 8\, c_m^{r\, 2}\, M_S^{r\, 2} \,
+\, 8 \, d_m^{r\, 2}\, M_P^{r\, 2} \, = \, -8\, \widetilde{\delta} \, , \label{eq:NLO_rel}
\end{eqnarray}
where the corrections
\begin{equation}
\delta_{_{\mathrm{NLO}}}^{(m)} \,=\, \frac{3 M_S^2}{32\pi^2F^2}
\left\{ 1 + \left(1-\frac{M_S^2}{M_P^2}\right)
\xi_{S\pi}^{(m)}  + 2\left(\frac{M_P^2}{M_S^2}-1\right) \xi_{P\pi}^{(m)} \right.
\left. - \frac{2
M_P^2}{M_S^2}\!\left(1-\frac{M_V^2}{M_A^2}\right) \xi_{V\pi}^{(m)}
\right\}
 \end{equation}
are known functions of the resonance masses:
\beqn
\xi_{S\pi}^{(1)} &= & 1 \,-\, \frac{6 M_S^2}{M_P^2}\, +\,
\left(\frac{4 M_S^2}{M_P^2}\,-\,\frac{6 M_S^4}{M_P^4}\right)
\ln{\left(\frac{M_P^2}{M_S^2}\,-\,1\right)}\, ,
\no\\
\xi_{P\pi}^{(1)} & = &  1\, +\,\frac{M_P^2}{M_S^2}
\ln{\left(1\,-\,\frac{M_S^2}{M_P^2}\right)} \,,\no
\\
\xi_{V\pi}^{(1)} & = & 1\, +\, \frac{3 M_V^2}{M_P^2}\left[
\frac{M_V^2}{M_P^2}\,-\,\frac{3}{2}\,+\,
\left(1\,-\,\frac{M_V^2}{M_P^2}\right)^2
\ln{\left(\frac{M_P^2}{M_V^2}\,-\,1\right)}\right]\,  ,
\no \\
\xi_{S\pi}^{(2)} & = & -4  \,+\, \left(2\,-\,\frac{4
M_S^2}{M_P^2}\right) \ln{\left(\frac{M_P^2}{M_S^2}\,-\,1\right)} \,,
\\
\xi_{P\pi}^{(2)} & = &  1 \,+\,
\ln{\left(\frac{M_P^2}{M_S^2}\,-\,1\right)} \,,
\no\\
\xi_{V\pi}^{(2)} & = &\frac{M_P^2}{M_S^2}\left(1\,-\, \ln \frac{M_S^2}{M_V^2}\right) \,-\, \frac{2M_V^2}{M_S^2} \left(1\,-\, \frac{M_V^2}{M_P^2} \right) \no \\ & &
 +\, \left(\frac{M_P^2}{M_S^2}\,+\,\frac{2M_V^2}{M_S^2} \right) \left(1\,-\,\frac{M_V^2}{M_P^2}\right)^2 \ln \left(\frac{M_P^2}{M_V^2}\,-\,1 \right)  .\no
\eeqn
Note that Eqs.~(\ref{eq:NLO_rel}) determine the NLO couplings $c_m^r$ and $d_m^r$:
\begin{eqnarray}
c_m^{r\,\,2}&=& \frac{F^2}{8} \frac{M_P^{r\,\,2}}{M_P^{r\,\,2}-M_S^{r\,\,2}} \left(1+\delta_{_{\rm NLO}}^{(1)}-\frac{M_S^2}{M_P^2}\delta_{_{\rm NLO}}^{(2)}-\frac{8}{M_P^2 F^2} \widetilde{\delta} \right) \, ,\nn \\
d_m^{r\,\,2}&=& \frac{F^2}{8} \frac{M_S^{r\,\,2}}{M_P^{r\,\,2}-M_S^{r\,\,2}} \left(1+\delta_{_{\rm NLO}}^{(1)}-\delta_{_{\rm NLO}}^{(2)} -\frac{8}{M_S^2 F^2} \widetilde{\delta}\right) \, .\label{dmr}
\end{eqnarray}

\section{$\mathbf{L_8^r(\mu)}$ at NLO} 

The low-momentum expansion of the R$\chi$T correlator in
Eq.~\eqn{eq:Pi_SP} reproduces the $U(3)_L\otimes U(3)_R$ $\chi$PT
result \eqn{eq:Pi_chpt}, with a definite prediction for the LEC
$L_8^r(\mu)$:
\beqn\label{eq:L_8_NLO}
\bar L_8^{U(3)}\!\!\!&& \equiv
\left[L_8^r(\mu) \,+\, \frac{\Gamma_8}{32\pi^2}\ln{\frac{\mu^2}{M_S^2}} \right]_{U(3)} \no \\ &&
 = \frac{F^2}{16} \left(\frac{1}{M_S^{r\, 2}}+\frac{1}{M_P^{r\,2}}\right)
\left\{ 1+  \delta_{_{\rm NLO}}^{(1)} - \frac{M_S^{r\, 2}\,\delta_{_{\rm NLO}}^{(2)}+8\,\widetilde{\delta}/F^2}{M_S^{r\, 2}+M_P^{r\, 2}}
\right\}
 - \frac{3\, \Delta}{256\pi^2}\, ,
\eeqn
with
\beqn \Delta & =& 1 -
\!\left(1-\frac{M_V^2}{M_A^2}\right)\!\left[
\frac{17}{6} -\frac{7M_V^2}{M_P^2} +\frac{4M_V^4}{M_P^4} + \!\left(
1-\frac{4M_V^2}{M_P^2} \right)\! \left( 1-\frac{M_V^2}{M_P^2}
\right)^2 \! \ln{\!\left(\frac{M_P^2}{M_V^2}-1\right)\!} \right]
 \no\\ & +
& \left(\frac{M_P^2}{M_S^2}-1\right) \left[ 2 +
\left(\frac{2M_P^2}{M_S^2}-1\right)
\ln{\left(1-\frac{M_S^2}{M_P^2}\right)}  +\frac{M_S^2}{6M_P^2} +
\frac{M_S^4}{M_P^4} -\frac{4M_S^6}{M_P^6}  \right] \no\\  & + &
\frac{M_S^4}{M_P^4} \left(1-\frac{M_S^2}{M_P^2}\right) \left(
3-\frac{4M_S^2}{M_P^2} \right)
  \ln{\left(\frac{M_P^2}{M_S^2}-1\right)} \, .
\eeqn
We have used the relations in Eqs.~\eqn{dmr} to eliminate the
explicit dependence on the effective couplings $c_m^r$ and $d_m^r$.

Eq.~\eqn{eq:L_8_NLO} modifies the large-$N_C$ result in
\eqn{eq:L_8_LO} with NLO corrections $\delta_{_{\mathrm{NLO}}}^{(1)}$, $\delta_{_{\rm NLO}}^{(2)}$ and
$\Delta$, which are fully known in terms of resonance masses. We have also taken into account the tiny correction $\widetilde{\delta}$.
Moreover, our calculation has generated the right
renormalization-scale dependence, giving rise to an absolute
prediction for the scale-independent parameter $\bar L_8^{U(3)}$.
Since we are working within the large-$N_C$ framework, the
Goldstone-nonet loops reproduce the non-analytic $\ln{(-t)}$
structure that arises in $U(3)_L\otimes U(3)_R$ \chpt. To make
contact with the usual $SU(3)_L\otimes SU(3)_R$ theory, we still
need to integrate out the singlet $\eta_1$ field. Computing the
massive one-loop $\eta_1$ contribution to $\Pi(t)$, one easily gets
the known relation \cite{KL:00} between the corresponding $L_8$
couplings in the two chiral effective theories. At $\cO(p^4)$ in the $U(3)_L \otimes U(3)_R$ theory, one finds:
\beqn
\bar L_8^{SU(3)} & = & \bar L_8^{U(3)}\, +\,
\frac{\Gamma_8^{SU(3)}-\Gamma_8^{U(3)}}{32\pi^2}
\ln{\frac{M_{\eta_1}^2}{M_S^2}}
 \,= \, \bar L_8^{U(3)}\, -\,
\frac{1}{384\pi^2}\ln{\frac{M_{\eta_1}^2}{M_S^2}}\, .\label{matching}
\eeqn
%

%
%

The different input parameters are defined in the chiral limit.
We take the ranges \cite{GL:85,PI:02,PDG,MVsplit,Kaiser} $M_V=(770\pm
5)$~MeV, 
$M_S^r=(1.14\pm 0.16)$~GeV,
$M_P^r=(1.3\pm 0.1)$~GeV, $M_{\eta_1}=(0.85 \pm 0.05)$~GeV
and $F=(89 \pm 2)$~MeV, and use Eq.~(\ref{log}) to fix $M_A$, keeping the constraint $M_P\geq M_S$ from Eqs.~(\ref{eq:cm_cd}) and imposing $M_A \geq 1$ GeV. The correction $\widetilde{\delta}$ turns out to be negligible. One obtains the numerical prediction
\bel{eq:L8bar}
\bar L_8^{SU(3)}\, =\, (0.4\,\pm\, 0.4)\cdot 10^{-3}\, .
\ee
The largest uncertainties originate in the badly known values of
$M_S^r$ and $M_P^r$, which already appear in the leading order
prediction \eqn{eq:L_8_LO}.
To account for the higher-mass intermediate states which have been neglected in
\eqn{eq:ImSS}, we have added an additional truncation error equal to
$0.12\cdot 10^{-3}$, the contribution of the heaviest included channel ($P\pi$). All
errors have been added in quadrature. At the usual \chpt\
renormalization scale $\mu_0=0.77$ GeV, Eq.~\eqn{eq:L8bar} implies
\be L_8^r(\mu_0)^{SU(3)} \,= \,(0.6\,\pm\, 0.4)\cdot 10^{-3}\, , \ee
to be compared with the value $L_8^r(\mu_0)^{SU(3)}=(0.9\pm 0.3)\cdot
10^{-3}$, usually adopted in $\cO(p^4)$ phenomenological analyses, or $L_8^r(\mu_0)^{SU(3)}=(0.62\pm 0.20)\cdot 10^{-3}$, obtained from  the $\cO(p^6)$ fit of Ref.~\cite{ChPT2loopsdbis}.

The sizable numerical difference between $\bar L_8^{SU(3)}$ and
$L_8^r(\mu_0)^{SU(3)}$ shows the large sensitivity of this coupling to
the \chpt\ renormalization scale. This is a general trend for those
LECs which are dominated by scalar or pseudoscalar resonance
exchanges. Therefore, to perform accurate phenomenological
applications one needs to control the renormalization scale
dependence, which requires a determination of the \chpt\ couplings
at NLO in $1/N_C$, like the one presented here for $L_8$.

\section{$\mathbf{C_{38}^r(\mu)}$ at NLO}

Following the same procedure used in Section~3, one is able to find a NLO prediction of the LEC $C_{38}^r (\mu)$:
\begin{eqnarray}
\bar C_{38}^{U(3)}\!\!\!&& \equiv
\left[C_{38}^r(\mu) \,-\, \frac{\Gamma^{(L)}_{38}}{32\pi^2} \ln{\frac{\mu^2}{M_S^2}} \right]_{U(3)} \!=\, \frac{F^4}{16 M_P^{r\,2}M_S^{r\,2}} \left( 1 + \frac{M_P^{r\,2}}{M_S^{r\,2}} + \frac{M_S^{r\,2}}{M_P^{r\,2}} \right) \times  \no \\
&& \times \left\{ 1\,+\,  \delta_{_{\rm NLO}}^{(1)} \, -\, \frac{(M_S^{r\,2}\, \delta_{_{\rm NLO}}^{(2)} + 8\, \widetilde{\delta}/F^2 )\,\left(M_S^{r\,2}+M_P^{r\,2}\right)}{M_P^{r\,4}+M_S^{r\,4}+M_P^{r\,2}M_S^{r\,2}} \right\} -\frac{9F^2\Delta'}{512 \pi^2 M_S^2} \,,
\end{eqnarray}
with
\begin{eqnarray}
\Delta' &&= 1+ \frac{1}{3}\left(\frac{M_P^2}{M_S^2}-1 \right) \left[ 6 -\frac{M_S^2}{M_P^2}+2 \left(\frac{3M_P^2}{M_S^2} - 2 \right) \ln \left( 1-\frac{M_S^2}{M_P^2} \right) \right] \no \\
&&
 +\frac{1}{3} \!\left(1-\frac{M_S^2}{M_P^2} \right)\! \left[ \frac{1}{6} + \frac{2M_S^2}{3M_P^2} + \frac{3M_S^4}{M_P^4} - \frac{10M_S^6}{M_P^6} 
 + \left(\frac{8M_S^6}{M_P^6}-\frac{10M_S^8}{M_P^8} \right)  \ln \left(\frac{M_P^2}{M_S^2} - 1 \right) \right] \no \\
&& +\frac{1}{3} \left(1-\frac{M_V^2}{M_A^2} \right) \Bigg[\frac{M_S^2}{2M_V^2}-\frac{28M_S^2}{3M_P^2}+\frac{19M_V^2M_S^2}{M_P^4}-\frac{10M_V^4M_S^2}{M_P^6}\no \\
&&
\qquad \qquad \quad \qquad + \left( \frac{4M_S^2}{M_P^2} -\frac{10M_V^2M_S^2}{M_P^4} \right) \left(1-\frac{M_V^2}{M_P^2} \right)^2 \ln \left( \frac{M_P^2}{M_V^2}-1 \right)  \Bigg]    \, .
\end{eqnarray}
We have used again Eqs.~(\ref{dmr}) to fix $c_m^r$ and $d_m^r$. As
in the case of $L_8^r(\mu)$, one has to make contact with the usual
$SU(3)_L \otimes SU(3)_R$ theory by computing the massive one-loop
$\eta_1$ contribution to $\Pi(t)$. It is straightforward to get the
expression that relates $\bar C_{38}$ in both theories:
\begin{eqnarray}
\bar C_{38}^{SU(3)} & = & \bar C_{38}^{U(3)}\, -\,
\frac{\Gamma_{38}^{(L)\,SU(3)}-\Gamma_{38}^{(L)\,U(3)}}{32\pi^2 }
\left( \ln{\frac{M_{\eta_1}^2}{M_S^2}}+\frac{1}{2}\right) -\frac{\Gamma_{8}^{SU(3)}-\Gamma_{8}^{U(3)}}{32\pi^2 }\frac{F^2}{2M_{\eta_1}^2}  \no \\
&=& \bar C_{38}^{U(3)}\, -\,
\frac{F^2}{192\pi^2}\left( \frac{1}{M_S^2} \ln{\frac{M_{\eta_1}^2}{M_S^2}}+\frac{1}{2M_S^2} -\frac{1}{4M_{\eta_1}^2} \right) \, , \label{c38inv}
\end{eqnarray}
where we have used the LO prediction of the $\cO(p^4)$ chiral
coupling $L_5=c_dc_m/M_S^2 = F^2/(4M_S^2)$~\cite{RChTa}.

Taking the same input parameters than in the previous section, one
gets the numerical prediction
\begin{eqnarray}
\bar C_{38}^{SU(3)} &=&\left( -1\,\pm\,6 \right)\cdot 10^{-6} \,.\label{c38invres}
\end{eqnarray}
At the usual $\chi PT$ renormalization scale $\mu_0 \,=\,0.77$ GeV, Eq.~(\ref{c38invres}) gives
\begin{eqnarray}
 C_{38}^{r}(\mu_0)^{SU(3)} &=&\left( 2\,\pm\,6 \right)\cdot 10^{-6}
 \, ,
\end{eqnarray}
showing again the large numerical sensitivity to the choice of
scale.

A sizable difference between the phenomenological value of some
$\cO(p^6)$ chiral couplings and their large-$N_C$ estimates was
pointed out in Ref.~\cite{difference}. Our result shows how the NLO
corrections in $1/N_C$ become relevant in some cases. These NLO
contributions must be considered for a proper determination of the
LECs. Our calculation reproduces the correct scale-dependence of
$C_{38}^r(\mu)$ at the NLO in $1/N_C$. However, the $\cO(p^6)$ LECs
contain additional dependences on $\mu$, which are suppressed by two
powers of $1/N_C$~\cite{ChPT2loops}. In order to remove the
corresponding numerical uncertainty, it would be necessary to
perform the calculation at the next-to-next-to-leading order in the
$1/N_C$ expansion.

\section{Conclusions}

The large--$N_C$ limit provides a solid theoretical framework to
understand the success of resonance saturation in low-energy
phenomenology \cite{PI:02}. However, this limit is unable to pin
down the scale dependence of the \chpt\ couplings. Although this is
a NLO effect in the $1/N_C$ expansion, its numerical impact is very
sizable in couplings which are dominated by scalar or pseudoscalar
exchanges.

In this paper we have presented a NLO prediction of the $\cO(p^4)$
coupling $L_8^r(\mu)$, which exactly reproduces its right
renormalization-scale dependence. Moreover, we have also determined
the $\cO(p^6)$ coupling $C_{38}^r(\mu)$ at the NLO, controlling its
$\mu$ dependence up to small NNLO effects.

The determination of this two LECs has been made possible through a
NLO calculation in $1/N_C$ of the $\Pi(t)\equiv
\Pi_{S}(t)-\Pi_{P}(t)$ correlator, in the chiral limit. We have used
the \rcht\ Lagrangian, within the SRA, to compute the one and
two-particle exchange contributions to the absorptive part of the
correlator. It is remarkable that, imposing a good short-distance
behaviour for the corresponding scalar and pseudoscalar spectral
functions, one fully determines the relevant contributing form
factors. Using a dispersion relation, we have reconstructed the
correlator, up to a subtraction term which has the same structure as
the tree-level one-particle contributions.

The stringent short-distance QCD constraints on $\Pi(t)$ have
allowed us to fix the subtraction constants in terms of resonance
masses, with the results shown in Eqs.~\eqn{dmr}. Therefore, we have
obtained a complete NLO determination of $\Pi(t)$, which only
depends on the pion decay constant $F$ and the meson masses $M_V$,
$M_A$, $M_S$ and $M_P$. Its low momentum expansion reproduces the
right \chpt\ expression, with explicit values for the LECs
$L_8^r(\mu)$ and $C_{38}^r(\mu)$.

Integrating out the $\eta_1$ field, one can further connect the
$U(3)_L\otimes U(3)_R$ and $SU(3)_L\otimes SU(3)_R$ effective
theories. This introduces an additional dependence on $M_{\eta_1}$.
At the usual scale $\mu_0=0.77$~GeV, we finally obtain the numerical
predictions:
\be L_8^r(\mu_0)^{SU(3)}\, =\, (0.6 \pm 0.4) \cdot 10^{-3}\, ,
\qquad\qquad C_{38}^r(\mu_0)^{SU(3)} \, =\, ( 2\pm 6) \cdot
10^{-6}\, . \ee

The ideas discussed in this article can be applied to generic Green
functions, which opens a way to investigate other chiral LECs at NLO
in the large-$N_C$ expansion. Further work along these lines is in
progress \cite{RSP:06}.

\section*{Acknowledgments}

We are indebted to J.~Portol\'es for many useful comments on the
topic of this work. I.R. is supported by a FPU contract of the
Spanish MEC. This work has been supported in part by China National
Natural Science Foundation under grants 10575002 and  10421503, by
the EU MRTN-CT-2006-035482 (FLAVIA{\it net}), by MEC (Spain) under
grant FPA2004-00996 and by Generalitat Valenciana under grants
ACOMP06/098 and GV05/015.

\appendix
\newcounter{ovidio}
\renewcommand{\thesection}{\Alph{ovidio}}
\renewcommand{\theequation}{\Alph{ovidio}.\arabic{equation}}
\renewcommand{\thetable}{\Alph{ovidio}.\arabic{table}}
\setcounter{ovidio}{1}
\setcounter{equation}{0}
\setcounter{table}{0}
\section{The one-loop correction $\Delta \Pi (t)|_\rho$}

In this appendix we show the explicit expression of the one-loop correction $\Delta \Pi(t)|_\rho$, generated by the considered two-particles exchanges, which has been calculated by using the dispersive method discussed in Section~2.

\begin{eqnarray}
\Delta \Pi_{_{S-P}} (t)|_{\eta\pi} & =& \frac{n_f}{2} \frac{1}{16 \pi^2}  \left(\frac{M_S^2}{M_S^2 - t}\right)^2  \left[ -1 + \frac{t}{M_S^2} - \ln{\left(\frac{-t}{M_S^2}\right)} \right] \,, \label{totc1} \\
\Delta \Pi_{_{S-P}} (t)|_{V\pi}& =&  \frac{n_f}{2}  \frac{1}{16 \pi^2}
\left( 1-\frac{M_V^2}{M_A^2} \right) 
\left(\frac{M_P^2}{M_P^2 - t}\right)^2 \left[ \left(1-\frac{t}{M_P^2}\right) \left(-\frac{ 2M_V^4}{t^2} - \frac{ 2 M_V^4}{t M_P^2} \right.  \right. \nn \\
& & \left. + \frac{5 M_V^2}{t} + 2 - \frac{ 9 M_V^2}{M_P^2} + \frac{ 6 M_V^4}{M_P^4} \right) +  2 \!\left( 1 -\frac{M_V^2}{t}\right)^3\!
\ln \!{\left(1-\frac{t}{M_V^2}\right)\!}
\nn \\
 &&\left.- 2\left( 1-\frac{4M_V^2}{M_P^2}+\frac{3M_V^2t}{M_P^4} \right)  \!\left(1-\frac{M_V^2}{M_P^2}\right)^2 \!  \! \ln \frac{M_P^2-M_V^2}{M_V^2} \!  
\! \right]\!   ,     \label{totc4}\\
\Delta \Pi_{_{S-P}} (t)|_{A\pi}& =&0\,, \phantom{\frac{1}{2}}\label{totc2}\\
\Delta \Pi_{_{S-P}} (t)|_{S\pi}& =&\frac{n_f}{2}\frac{1}{16 \pi^2}   \left( 1- \frac{M_S^2}{M_P^2} \right) 
\left(\frac{M_P^2}{M_P^2 - t}\right)^2 
\left\{-\frac{2M_S^4}{t^2}+\frac{M_S^2}{t}+\frac{8M_S^4}{M_P^4}-\frac{2M_S^2}{M_P^2} \right. \nn \\
&&+\frac{M_S^2t}{M_P^4}-\frac{6M_S^4t}{M_P^6} + \frac{2M_S^4}{t^2}\left(1-\frac{M_S^2}{t}\right)  \ln{\left(1-\frac{t}{M_S^2}\right)}  \nn \\
&&\left.+\frac{2M_S^4}{M_P^4}\left( 3-\frac{4M_S^2}{M_P^2}-\frac{2t}{M_P^2}+\frac{3M_S^2t}{M_P^4} \right)  \ln \frac{M_S^2}{M_P^2-M_S^2} \right\}\, , \label{totc5}\\
%
\Delta \Pi_{_{S-P}} (t)|_{P\pi}& =& \frac{n_f}{2} \frac{1}{16 \pi^2}
\frac{M_S^2 (M_P^2-M_S^2)}{(M_S^2 - t)^2}
 \left[ -2 +\frac{2t}{M_S^2} -  2\left( 1 - \frac{M_P^2}{t}\right)  \ln{\left(1-\frac{t}{M_P^2}\right)}\nn\right. \\
&& \left. + 2\left( 1-\!\frac{2 M_P^2}{M_S^2} + \frac{M_P^2 t}{M_S^4} \right)  \ln\frac{M_P^2-M_S^2}{M_P^2}  \right] . \label{totc3}
\end{eqnarray}


\end{document}